\documentclass{pasj00}
\draft

\begin{document}
\SetRunningHead{Dogiel et al.}{6.4 keV Line Emission from
Molecular Clouds in the Galactic Center}
\Received{2000/12/31}
\Accepted{2001/01/01}

\title{Line and Continuum Emission from the Galactic Center.
\texttt{III. Origin of 6.4 keV Line Emission from Molecular Clouds
in the Galactic Center}}

\author{Vladimir \textsc{Dogiel}$^{1,2}$,  Kwong-Sang
\textsc{Cheng}$^3$, Dmitrii {\sc Chernyshov}$^{2,4}$, Aya {\sc
Bamba}$^{1}$, Atsushi {\sc Ichimura}$^{1}$, Hajime {\sc
Inoue}$^{1}$, Chung-Ming {\sc Ko}$^5$, Motohide {\sc
Kokubun}$^{1}$, Yoshitomo {\sc Maeda}$^{1}$, Kazuhisa {\sc
Mitsuda}$^{1}$, and Noriko Y. {\sc Yamasaki}$^1$}

\affil{$^1$Institute of Space and Astronautical Science, 3-1-1,
Yoshinodai, Sagamihara, Kanagawa, 229-8510, Japan}
  \affil{$^2$P.N.Lebedev Institute, Leninskii
pr, 53, 119991 Moscow, Russia, dogiel@lpi.ru}
 \affil{$^3$Department of
Physics, University of Hong Kong, Pokfulam Road, Hong Kong, China}
\affil{$^4$Moscow Institute of Physics and Technology, Institutskii lane, 141700 
Moscow Region, Dolgoprudnii, Russia} \affil{$^5$Institute of
Astronomy, National Central University, Jhongli 320, Taiwan}

%

\KeyWords{Galaxy: center - X-rays: diffuse background -  ISM: molecular clouds: cosmic rays} 

\maketitle

\begin{abstract}

We analyze the 6.4 keV line and continuum emission from the
molecular cloud Sgr B2 and the source HESS J1745-303, which is
supposed to be a complex of molecular gas. From the HESS results
it follows that Sgr A$^\ast$ is a source of high energy protons,
which penetrate into molecular clouds producing there a TeV
gamma-ray flux. We present arguments that Sgr A$^\ast$ may also
produce a flux of subrelativistic protons which generate the 6.4
keV line and bremsstrahlung continuum emission from the clouds.

\end{abstract}

\section{Introduction}
In the two previous papers of this series  (Dogiel 2009ab) we
presented arguments in favor of the injection of subrelativistic
protons by star accretion on the central black hole. We showed
that this process might explain the origin of hot plasma in the
Galactic center and could produce a flux de-excitation gamma-ray
lines and  non-thermal X-ray emission
 observed by {\it Suzaku} in the range 14 to 40 keV
\citep{yuasa}. Below we analyze the origin of the 6.4 keV emission
from molecular clouds in the Galactic center (GC) which may also
be generated by  these  subrelativistic protons.

One of the first observations of X-ray emission from molecular
clouds was performed by \citet{suny} who found a flux of X-rays
from compact sources in the GC. They assumed that a large portion
of this flux arises from Thomson scattering (reflection) by dense
molecular clouds which are irradiated by a nearby X-ray source,
e.g., by a flux from the central supermassive black hole, which
was active in the recent past ($\sim 300-400$ years ago) but is
almost unseen at present. They predicted also a bright fluorescent
K$\alpha$ line in the scattered spectrum of the clouds due to the
K-absorption of photons with energies $E>7.1$ keV. This line was
discovered then with the ASCA telescope from the molecular cloud
Sgr B2 \citep{koya1,mura} and from Sgr C \citep{mura2}. Later on
6.4 keV emission was discovered also in other molecular clouds
(\cite{bamba02,predehl,nobuk,naka,koya09}). In subsequent
publications based on observations with {\it Chandra}
\citep{mura1} and    {\it Suzaku} \citep{koya3, naka}
 arguments were presented that the Sgr B2 and Sgr C clouds are, indeed,  X-ray reflection
nebulas (XRNs) irradiated by Sgr A$^{\ast}$ which was X-ray bright
300 years ago.  \citet{mura03} analyzing intensities of 6.4 keV
line emissions measured by {\it Chandra} from the giant molecular
clouds in the GC region: Sgr B2, Sgr C, and M0.11-0.08, obtained
the luminosity history of the Galactic nuclei Sgr A$^\ast$ during
the last 500 years. They concluded that Sgr A$^\ast$ was as
luminous as $10^{39}$ erg s$^{-1}$ a few hundreds years ago, and
has dimmed gradually since then.

\citet{suny1} provided theoretical treatments of continuum and
line emission from molecular clouds exposed by external sources.
They showed that a short time variability was the key to
investigate the nature of radiating sources from the shape and
time variations of the 6.4 keV line. Since the light crossing time
of Sgr B2 is about 30 years, they predicted a decline of the 6.4
keV line flux by a factor of 2 for the period of 10 years.

Using archival data of {\it ASCA}, {\it BeppoSAX}, {\it Chandra},
and {\it XMM} observatories \citet{rev04} found no significant
variability of the line flux from Sgr B2 during the period 1993 -
2001. The constancy of the line flux  meant that the luminosity of
Sgr A$^\ast$ remained approximately constant for more than 10
years a few hundred years ago, while the fact that other molecular
clouds in the GC region also shined in the 6.4 keV line indicated
that the entire period of activity lasted much longer than 10
years.

Very recently \citet{koya4} and \citet{inui} presented new {\it
Suzaku} data which, in combination with {\it ASCA, XMM-Newton} and
{\it Chandra} data, showed a time-variability of the 6.4 keV line
emission from Sgr B2.  It was concluded  from a cross-correlation
between the X-ray telescopes which measured this flux during
different periods of time that this line exhibited a brightest
peak in 2000 but fell down to 60\% of the peak in 2005.

Since in the framework of XRN model the 6.4 keV line emission and
the reflected continuum fluxes from molecular clouds are generated
by the same primary X-ray radiation from an external source one
expects a time-correlation between these fluxes. However, INTEGRAL
observations indicate that the continuum $18 - 60$ keV flux was
constant within 25\% during $2003 - 2004$ \citep{rev04}. However,
6 year monitoring of the Sgr B2 region by {\it INTEGRAL}
\citet{terr09} gave evidence that a continuum flux from there in
the range 20-40 keV was also time-variable with a decrease of more
than 30\% .

 Recently  \citet{bamba} found  6.4 keV emission
 in the direction of  the  source of
TeV gamma-rays (HESS J1745-303). This source is supposed to be
 a complex of molecular gas. \citet{aha}  showed that the TeV
 gamma-ray emission is most probably produced by a flux of high
 energy protons. From the spectral analysis it was concluded  that
this complex might also be an XRN source irradiated by Sgr
A$^\ast$ or the nearby SNR G359.1-05.

 If this
 molecular cloud is filled with high energy charged protons, as
 follows from the HESS data, then the
 question is whether these particles can also generate emission of
 the iron line from this cloud. We have reasons to assume that Sgr
 A$^\ast$ produces high energy protons not only in the TeV energy range
 but also in the GeV (see \cite{cheng2}) and the MeV (see
 \cite{dog}) ranges. If so, then the XRN interpretation cannot be
 considered as unique and others are also possible, especially as
 some results of observations of the 6.4 keV line emission from clouds
 do not fall into the XRN model, that cannot be completely ignored.

Thus, \citet{predehl} presented their measurements of the 6.4 keV
line from the GC made with {\it XMM-Newton}. They measured this
emission from several molecular complexes situated at distances of
30 pc to 115 pc from Sgr A$^{\ast}$ and found that their surface
brightness did not differ very much despite of their different
distances from Sgr A$^\ast$, though one could expect that more
distant filaments should be dimmer at least by a factor of ten if
they are XRNs. A key characteristic of the XRN model is a
pronounced K-absorption edge at 7.1 keV in clouds. However, the
{\it XMM-Newton} measurements are completely consistent with
interstellar absorption only and do not show a significant
absorption from clouds.

An alternative interpretation of the origin of the K$\alpha$
fluorescent line in the Galaxy is its excitation by
subrelativistic charged particles, i.e. electrons or protons (see
 \cite{dog4,val,yus1,yus2,yus3}). Below we define this model as sources of X-rays
 emitted by charged particles (XECP).

We derive characteristics of the line and continuum emission from
the clouds Sgr B2 and the source HESS J1745-303 assuming that this
emission is generated by a flux of  protons produced by Sgr
A$^\ast$ from star accretion onto the central black hole.

\section{Medium  Parameters}
The medium parameters  near the GC are quite uncertain. Even the
total mass of the most  massive cloud Sgr B is poorly known. The
estimated mass in the 42 pc diameter ranges from $2\cdot 10^5$ to
$7\cdot 10^6$ M$_\odot$ \citep{oka}.  The cloud optical depth to
Thomson scattering of X-ray photons ($\tau \sim
\sigma_Tn_{H_2}r$), where $n_{H_2}$ is the number density of gas
and $r$ is the cloud radius, is estimated to have the value
$\tau\simeq 0.4$ \citep{rev04}.

The total mass of the complex HESS J1745-303 is estimated by the
value $5\cdot 10^4$ M$_\odot$ \citep{aha}. In Table \ref{bamba}
taken from \citet{bamba} we present the total gas mass, the
angular and linear sizes of  region emitting the 6.4 keV line and
the projection distances from Sgr A$^{\ast}$ for these two clouds.
\begin{table}[h]
\caption{Characteristics of molecular clouds.}
\centering
\begin{tabular}{|p{3.3cm}p{2.0cm}p{2.0cm}|}
 \hline
&Sgr B2& J1745-303 \\
\hline &&\\
M[M$_\odot$]&$6\cdot 10^6$&$5\cdot 10^4$\\
&&\\
Angular size[deg.]&0.05&0.3\\
Linear size[pc]&7&40\\
&&\\
Distance[deg.]&0.7&1.2\\
Linear distance[pc]&$\sim 100$&$\sim 200$\\
 \hline
\end{tabular}\label{bamba}
\end{table}
The angular distances give only projection values, therefore,  we
estimate linear distances from Sgr A$^\ast$ to the clouds in the
framework of XECP model.

The iron abundance is also poorly known. Direct estimations of
this value provided by the {\it Suzaku} group \citep{koya2,koya09}
gave the iron abundance in the intercloud medium at the GC from 1
to 3.5 solar. \citet{rev04} got the iron abundance for the cloud
Sgr B2 at about 1.9 solar.

The average plasma density of the intercloud medium  in the GC
ranges between $n\simeq 0.1$ - $0.4$ cm$^{-3}$. The plasma
temperature is around $T\simeq 6.5$ keV (see
\cite{koya1,muno,koya2}).

With all these uncertainties of parameters it is difficult to get
reliable quantitative estimates. We can conclude only whether the
XECP model can in principle reproduce these set of observational
data or not. More reliable quantitative conclusions can be derived
from future experiments.

\section{Flux Parameters}

 Since molecular clouds are extended sources of 6.4 keV and
continuum emission and their boundaries are not clear, their
fluxes estimated from observation  depend  on the size of the
source and background regions from which the spectrum is
extracted. This circumstance complicates significantly estimations
of emission fluxes from molecular clouds.

From observations the following parameters of the continuum in the
range 2(4)-10 keV range and 6.4 keV line emission  were obtained
for Sgr B2 and HESS J1745-303 \citep{koya1,mura,mura1,koya3, aha,
bamba,inui}.

They are presented in Tables \ref{t6.4} and \ref{bamba1}. We make
absorption corrections of the {\it Suzaku} upper limit
\citep{bamba} and the {\it XMM-Newton} upper limit \citep{aha} of
the continuum emission from HESS J1745-303  for the average gas
column density $L_H^{av}\simeq 1.3\cdot 10^{22}$ cm$^{-2}$
 and for the extreme value of
$L_H^{ex}\simeq 4.6\cdot 10^{23}$ cm$^{-2}$ which follows from the
\citet{aha} estimation of the cloud gas density equaling $5\cdot
10^3$ cm$^{-3}$. These values are shown in Table \ref{bamba1},
without and with brackets respectively.

\begin{table}[h]
\caption{6.4 keV Line flux  as observed from Earth ($F_{6.4}$) and
 the continuum emission in the
range 2(4)-10 keV ($\Phi_{2(4)-10~{\footnotesize \mbox{keV}}}$)
from Sgr B2$^a$.} \centering
\begin{tabular}{|p{1.8cm}p{2.5cm}p{3.0cm}|}
 \hline
Telescope&$10^{5}\cdot F_{6.4}$&$10^{-33}\cdot\Phi_{2(4)-10~{\footnotesize \mbox{keV}}}$\\
&(ph
cm$^{-2}$s$^{-1}$)&( erg s$^{-1}$) \\
&&\\
\hline
ASCA&$16.3$&$110- 140 $\\
\hline
Chandra&13.7-17.1&$80-120$\\
\hline Suzaku&11.4&97\\
 \hline
\end{tabular}\label{t6.4}
\begin{list}{}{}
\item\tiny{$^{\mathrm{a}}$ Absorption corrected luminosity. The intensities of 6.4 keV
line were taken from Tables 2 and 3 of \citet{inui}.}
\end{list}
\end{table}

\begin{table}[h]
\caption{Flux of 6.4 keV  line as observed from Earth,
($F_{6.4}$),  and  upper limits of the continuum emission in the
range 2-10 keV ($\Phi_{2-10~{\footnotesize \mbox{keV}}}$) from
HESS J1745-303$^a$.} \centering
\begin{tabular}{|p{1.6cm}p{2.7cm}p{3.2cm}|}
 \hline
Telescope&$10^{5}\cdot F_{6.4}$
&$10^{-33}\cdot\Phi_{2-10~{\footnotesize \mbox{keV}}}$  \\
&(ph cm$^{-2}$s$^{-1}$)&(erg s$^{-1}$)\\
&&\\
\hline Suzaku&1.1&$<3.9~(<80)$\\
&&90\% confidence limit\\
\hline
XMM&-&$<9.6~(<226)$\\
-Newton&&99\% confidence limit\\
 \hline
\end{tabular}\label{bamba1}
\begin{list}{}{}
\item\tiny{$^{\mathrm{a}}$ Absorption corrected luminosity.}
\end{list}
\end{table}

Below we analyze whether this emission can be generated by
subrelativistic protons.

\section{Spectrum of Subrelativistic Protons in the GC
 Region}

In order to calculate a flux of X-ray emission from  clouds
produced by protons we should estimate their density nearby these
clouds and inside them. The average diffusion coefficient of
cosmic rays in the Galactic disk equals $D\simeq 10^{27}$
cm$^2$s$^{-1}$ \citep{ber90}. Each event of star accretion on the
central black hole produces about $Q=10^{57}$ subrelativistic
protons with an energy of about $E_m\sim 100$ MeV \citep{dog}. The
characteristic time of one solar mass star capture is
$\tau_k\simeq 10^{4}$ years \citep{syer}. The injection function
for a single star capture event can be presented  as a simple
Gaussian distribution

\begin{equation}\label{Qesc}
        Q_k(E)=\frac{N}{\sigma\sqrt{2\pi}}
        \exp\left[-\,\frac{(E-E_{m})^2}{2\sigma^2}\right]\,,
\end{equation}
where we take the width $\sigma=0.03E_{m}$ with $E_{m}=100$ MeV,
and $Q$ is total amount of particles ejected by one stellar
capture event. The index $k$ denotes the number of capture events.
Then time of the $k$th capture is $t_k=k\times \tau_k$.

The  distribution of MeV protons in the GC region can be
calculated from the three-dimensional diffusion equation:
\begin{equation}\label{pr_state}
 \frac{\partial N}{\partial t}+\frac{\partial}{\partial E}\left( b(E) N\right) - \nabla D\nabla N
 = Q(E,t)\,,
\end{equation}
where the rate of ionization losses is
\begin{equation}
\left(\frac{dE}{dt}\right)_i\equiv b(E)=-\frac{2\pi n
e^4\ln\Lambda}{mv}\,.
\end{equation}
Here $m$ is the electron mass, $v$ is the proton velocity and
$\ln\Lambda$ is the Coulomb logarithm. The source term has the
form
\begin{equation} Q(E,
{\bf r}, t) = \sum \limits_{k=0}Q_k(E)\delta(t - t_k)\delta({\bf
r})\,,
\end{equation}
where $t_k$ is the injection time, and the functions $Q_k(E)$ are
given by Eq. (\ref{Qesc}). The solution of Eq. (\ref{pr_state}) is
presented in \citet{dog}.

\begin{eqnarray}
N({\bf
r},E,t)=&&\sum\limits_{k=0}\frac{Q_k\sqrt{E}}{\sigma\sqrt{2\pi}Y_k^{1/3}}\theta(t-t_k)
\times\nonumber\\
&&\times\frac{\exp\left[-\frac{\left(E_{esc}-Y_k^{2/3}\right)^2}{2\sigma^2}-\frac{{\bf
r}^2}{4D(t-t_k)}\right]}{ \left(4\pi D(t-t_k)\right)^{3/2}}\,,
\label{sol1}
\end{eqnarray}
where $\theta(t-t_k)$ is the Heaviside (step) function, and
\begin{equation}
Y_k(t,E)=\left[\frac{3a}{2}(t-t_k)+E^{3/2}\right]\,.
\end{equation}

In Table \ref{t2} we summarize the parameters of proton production
and propagation in the intercloud medium of the GC, which we use
for calculations.
\begin{table}[h]
 \caption{Parameters of proton production and
propagation in the hot GC plasma} \centering
\begin{tabular}{|p{1.0cm}|p{1.1cm}|p{1.1cm}|p{1.0cm}|p{1.0cm}|p{1.0cm}|}
 \hline
Part.inj. numb.&Star capt. time
&Part.inj. energy&Plasma dens.&Plasma temp.&Diff. coeff.\\
\hline
$N_p$&$\tau_k$&$E_m$&n&T&D\\
\hline
$10^{57}$ prot.&$10^4$ year&$100$ MeV&0.2 cm$^{-3}$&6.5 keV&$ 10^{27}$ cm$^2$s$^{-1}$\\
\hline
\end{tabular}\label{t2}
\end{table}

\section{Proton density inside molecular clouds}
We denote by $N_c(E)$ the  proton spectrum on the surface of
clouds  which is calculated from Eq. (\ref{sol1}).
 The proton distribution inside the cloud depends on the processes
of proton penetration into dense neutral gas. This mechanism is
also uncertain but as it was shown in \citet{dog87} that strong
fluctuations of the magnetic field  induced by the observed gas
turbulence inside molecular clouds made particle propagation there
similar to diffusion with the  coefficient $D_c$ ranging from
$10^{24}$ to $10^{26}$ cm$^2$s$^{-1}$ depending on cloud
parameters. In the framework of the one-dimensional diffusion
approximation we calculate the proton distribution inside
molecular clouds.

In order to calculate the distribution of protons inside molecular
clouds we should derive their spectrum at cloud surfaces. The
linear distances from Sgr A$^\ast$ to the clouds was derived in
order to reproduce the line $F_{6.4}$ and continuum $\Phi_x$
emission from the clouds. These distances was estimated as $\sim
100$ pc for Sgr B2 and $\sim 450$ pc for HESS J1745-303.

\begin{figure}[h]
  \begin{center}
    \FigureFile(110mm,80mm){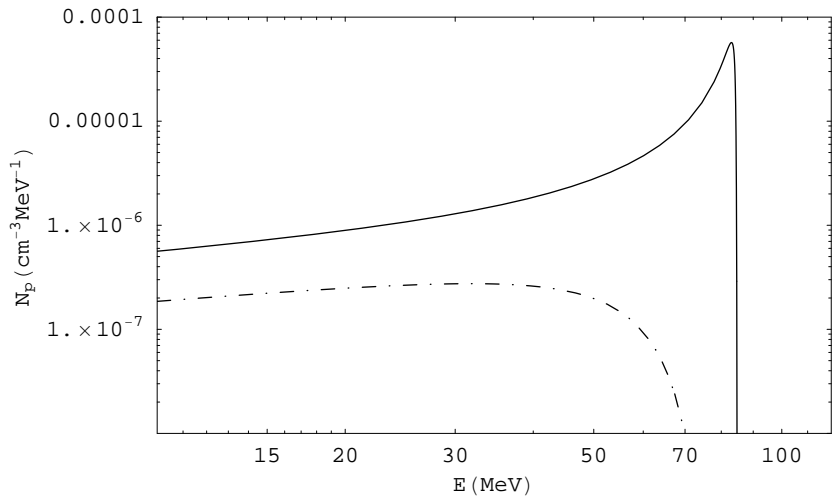}
  \end{center}
  \caption{Spectra of protons injected by Sgr A$^\ast$ near the clouds Sgr B2
  (solid line) and  HESS J1745-303 (dashed-dotted line). The distance from Sgr A$^\ast$
  to the cloud Sgr B2 is 100 pc and to HESS J1745-303 - 450 pc. These values were chosen
   in order to reproduce the observational data from Table \ref{t6.4}-\ref{bamba1}.}
  \label{p_out}
\end{figure}

Spectra of subrelativistic protons $N_c(E)$ calculated from
Eq.(\ref{sol1}) for the clouds Sgr B2 (solid line) and HESS
J1745-303 (dashed-dotted line) are shown in Fig. \ref{p_out}.

The gas distribution in the cloud Sgr B2 is highly nonuniform. It
has an envelope that extends to $\sim 45$ pc with the average
density $\sim 10^3$ cm$^{-3}$ and a dense core of 5-10 pc with the
density about $\sim 10^6$ cm$^{-3}$. Its density distribution can
be described as  (see \cite{lisg,lisg1})
\begin{equation}
\left(\frac{n_{H_2}}{1~\mbox{cm$^{-3}$}}\right)=5.5\times 10^4
\left(\frac{r}{1.25~\mbox{pc}}\right)^{-2}+2.2\times 10^3\,.
\end{equation}
For calculations we take the average value of the gas density
there to be $\bar{n}_{H_2}=10^4$ cm$^{-3}$. Almost the same gas
density for HESS J1745-303 was estimated by \citet{aha}.

The iron abundance inside these clouds were taken to be 2 solar,
though similar calculations can be performed for other set of
parameters. Spatial parameters of clouds, which we used for
calculations,  are shown in Table \ref{clouds}.
\begin{table}[h]
 \caption{Molecular cloud parameters (denoted by "c")}
 \centering
\begin{tabular}{|p{1.5cm}|p{1.6cm}|p{1.6cm}|p{1.9cm}|}
 \hline
Gas dens.&Gas temp.&Fe Abund.&Diff. coeff.\\
\hline
$\bar{n}_c$&$T_c$&$\eta_c$&$D_c$\\
\hline
$10^4$ cm$^{-3}$&100 eV&$7.4\cdot 10^{-5}$&$10^{25}$ cm$^2$s$^{-1}$\\
\hline
\end{tabular}\label{clouds}
\end{table}

The spectrum of protons, $\tilde{N}_p(E,x)$, penetrating into the
molecular clouds can be calculated in the framework of
one-dimensional diffusion equations for the derived spectrum
$N_c(E)$ on cloud surfaces. Here $x$ is the coordinate from the
cloud surface  to the cloud center.
\begin{equation}\label{pr_cl}
 \frac{\partial}{\partial E}\left( b_c(E) \tilde{N}\right) -  D_c\frac{\partial^2}{\partial x^2} \tilde{N}
 = 0\,,
\end{equation}
with the boundary conditions
\begin{equation}
\tilde{N}|_{x=0}=N_c,~~~~~~~~~~~~\tilde{N}_p|_{x=\infty}=0\,.
\end{equation}

The solution of the equation can be obtained by the method of
images \citep{morse}:
\begin{eqnarray}
\tilde{N}(E,x)=&&\frac{x}{\mid b_c(E)\mid}\int\limits_E^{E_m}
\frac{dE_0
N_c(E_0)}{\sqrt{D_c}\cdot\tau_c(E,E_0)^{3/2}}\times\nonumber\\
&&\times\exp\left[-\frac{x^2}{4D_c\cdot\tau_c(E,E_0)}\right]\,.
\label{n_cloud}
\end{eqnarray}
Here $b_c(E)$ is the rate of ionization losses inside the cloud
and
\begin{equation}
\tau_c(E,E_0)=\int\limits_{E_0}^E\frac{dt}{b_c(t)}\,.
\end{equation}

As follows from Eq. (\ref{n_cloud}),  subrelativistic protons are
unable to fill the whole volume of the cloud. They penetrate into
the clouds from outside for a depth $\sim\sqrt{\tau_c(E)D_c}$,
which is about $0.1-0.3$ pc.

\section{Continuum and 6.4 keV fluxes from molecular clouds}

In order to calculate the flux of 6.4 keV line  from the clouds we
used the proton distribution in the cloud (\ref{n_cloud}) and the
cross-section of $K\alpha$ vacancy production, $\sigma_K$, by
protons from \citet{garcia}.

Then the flux of 6.4 keV line is calculated from
\begin{equation}
F_{6.4}=\frac{ R^2\eta\omega_K \bar{n}_c}{d^2}\int\limits_0^\infty
dx\int\limits_E v(E)\sigma_K \tilde{N}(E,x)dE\,,
\end{equation}
where $\omega_K$ is the fluorescence yield of X-ray photon
emission, which is about 0.3 for iron, and $\eta$ is the iron
abundance, $R$ is the  radius of the region emitting the 6.4 keV
line (see Table \ref{bamba}).

The Doppler line width generated by proton impact is  broader than
that produced by e.g., electron impact (see \cite{dog4}). For the
width of the iron line observed by  the {\it Suzaku} the velocity
of the iron atoms cannot be higher than 150 km s$^{-1}$
(\cite{koya2}). The average velocity of iron atoms after collision
with a subrelativistic proton is  smaller than 100 km s$^{-1}$.
Therefore, for protons with energies $E<100$ MeV the width line is
about $\lesssim 30$ eV that is inside  the {\it Suzaku} Gaussian
of the 6.4 keV line width $\sim 120$ eV \citep{ebi2}.

Protons penetrating into molecular clouds generate also X-ray
continuum by inverse bremsstrahlung. The cross-section of inverse
bremsstrahlung is \citep{haya}
\begin{equation}
  {{d\sigma_{br}\over{dE_x}}}={8\over 3}{Z^2}{{e^2}\over{\hbar c}}\left({{e^2}
  \over{m{c^2}}}\right)^2{{m{c^2}}\over{E^\prime}}{1\over{E_x}}
\ln{{\left(\sqrt{E^\prime}+\sqrt{{E^\prime}-{E_x}}\right)^2}\over{E_x}}\,.
\label{sbr}
\end{equation}
Here  $E^\prime = (m/M)E_p$. Then the flux of inverse
bremsstrahlung radiation can be calculated from
\begin{equation}
\Phi_x=4\pi R^2\int\limits_{0}^\infty N_p(E,{\bf
r},t){{d\sigma_{br}\over{dE_x}}}\mathrm{v}_pn~dx\,.
\end{equation}

The results of the flux calculations are summarized in Table
\ref{X_clouds}.
\begin{table}[h]
\caption{Fluxes of 6.4 keV Line ($F_{6.4}$) as observed from Earth
and the total flux of proton inverse bremsstrahlung in the range
2-10 keV ($\Phi_{2-10~\mbox{\footnotesize keV}}$) from Sgr B2 and
HESS J1745-303.} \centering
\begin{tabular}{|p{1.8cm}p{2.2cm}p{3.3cm}|}
 \hline
Cloud&$10^{5}\cdot F_{6.4}$ &$10^{-33}\cdot\Phi_{2-10~\mbox{\footnotesize keV}} $ \\
&(ph cm$^{-2}$s$^{-1}$)&(erg s$^{-1}$)\\
 \hline
SGR B2&11&80\\
\hline
J1745-303&1.1&7.6\\
\hline
\end{tabular}\label{X_clouds}
\end{table}

From a comparison of the calculation results  with the data for
the clouds Sgr B2 and HESS J1745-303 (see Tables \ref{t6.4},
\ref{bamba1} and \ref{X_clouds}) we see their coincidence. This
means that the XECP model of X-ray production by protons inside
molecular clouds is reasonable.

\begin{figure}[h]
  \begin{center}
    \FigureFile(110mm,80mm){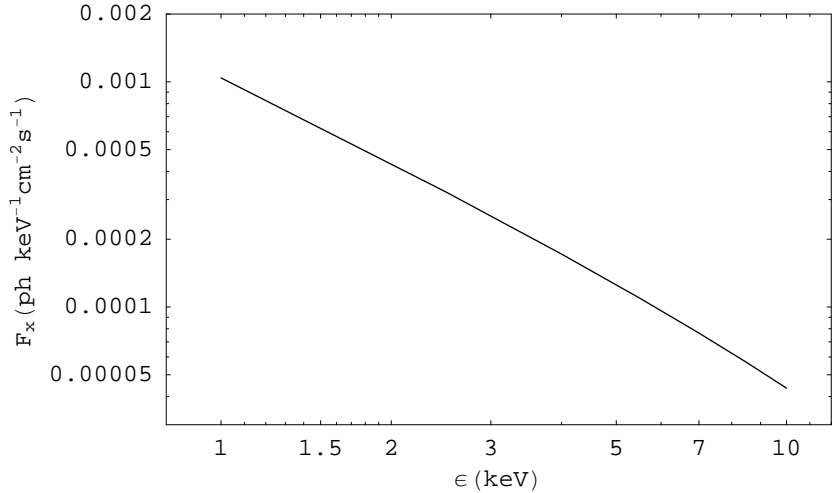}
  \end{center}
  \caption{Spectrum of inverse bremsstrahlung radiation from Sgr B2 as observed at Earth.}
  \label{inbr}
\end{figure}

The spectrum of bremsstrahlung photons from  the cloud Sgr B2 as
observed from Earth is shown in Fig. \ref{inbr}.

We notice that the $K\alpha$ vacancies and bremsstrahlung emission
can also be produced  by primary electrons generated by accretion
processes and by knock-on electrons generated directly inside
clouds by collisions of primary protons with the gas. However
estimates show that the energy of primary electrons is about 50
keV for the used accretion parameters. Therefore, their lifetime
is relatively short because of ionization losses.  The
contribution of secondary electrons to the total X-ray flux is
about three times smaller than that of primary protons
\citep{dog4}.

 Our analysis shows that there should a component of
nonthermal continuum X-ray emission whose intensity is
proportional to the intensity of the 6.4 keV line, and as follows
from \citet{dog_pasj} the intensity of the 6.7 keV line and
nonthermal continuum from the plasma should also correlate with
each other. This conclusion naturally explains the results
presented by \citet{koya09} who found two components of nonthermal
emission from the GC whose intensities are proportional to the
fluxes of the 6.7 and 6.4 keV lines, respectively.  Both
components have the same spectral index, that is not surprising in
the framework of XEPC model because emission of these components
from the hot plasma and the cold gas in the GC is generated by the
same flux of protons.

Since the total mass of molecular gas in the inner Galaxy can be
as high as $7\cdot 10^7M_\odot$, i.e about one order of magnitude
larger than the mass of Sgr B2, then molecular clouds may
contribute a significant part of the total hard X-ray flux from
the GC.

\section{Heating and ionization of molecular gas in the GC by
subrelativistic protons}

Ionization and heating of molecular clouds in the Galactic disk
 is an old problem (for  reviews on this subject (see \cite{dalmc},
 and \cite{spj}). We still do not know the
unobserved energetic radiation
 which maintains the heating and ionization state of the interstellar gas.
Hypothetical sources which in principle could deposit  significant
power into the interstellar gas were considered  to be either soft
X-rays \citep{siwer} or a flux of  cosmic rays \citep{haya1,spt}
in the form of protons with energies $2 - 10$ MeV (see  e.g.
\cite{nath} and \cite{indri}) or in the form of subrelativistic
electrons (see e.g. \cite{saher} and \cite{dog02}).

Parameters of the molecular gas in the GC are even more specific.
As \citet{yus3} noticed the molecular gas in GC is heated up to a
temperature higher than in the disk, $T_c\sim 100-200$ K .
Therefore, a global heating mechanism is needed there to explain
the high gas temperature.

\citet{yus1,yus2}  assumed that the processes of  6.4 keV line
emission and the gas ionization and heating are produced by
low-energy cosmic ray electrons with energies below 1 MeV which
are completely absorbed by molecular clouds. The inferred
ionization rates $\zeta$ of the GC clouds based on the 6.4 keV
line measurements range between $2\cdot 10^{-14}$ s$^{-1}$H$^{-1}$
and $5\cdot 10^{-13}$ s$^{-1}$H$^{-1}$.  Calculations of
\citet{neuf} showed that  the ionization rate $\zeta$ of the order
of $ 10^{-13}$ s$^{-1}$H$^{-1}$  is high enough in order to heat
the molecular gas in the GC up to 200 K.

In the framework of the XECP model the density of subrelativistic
protons in the GC is almost uniform within 50 pc that explains: a)
an almost uniform temperature distribution there  (see, e.g.,
\cite{nobuk}); and b) a constant surface brightness of the
molecular clouds in the GC, which does not differ very much
despite of their different distances of molecular clouds from Sgr
A$^\ast$ \citep{predehl}. If the 6.4 keV flux is provided by
protons, then in the central region we expect, indeed, a constant
emissivity of 6.4 keV line in clouds independently of the distance
from Sgr A$^\ast$.

We estimate the ionization rate $\zeta$ per one atom of H, which
can be provided by subrelativistic protons in Sgr B2 from the
relation

\begin{equation}
\zeta\simeq {\int\limits_E}\sigma_i v
{{dN}\over{dE}}dE~\mbox{(sec$^{-1}$H$^{-1}$)}\,.
\end{equation}
where the ionization cross-section $\sigma_i$ was taken from e.g.
\citet{spi:68}.

As calculations show the rate of ionization in  clouds is strongly
nonuniform. It is very high at a cloud surface,  $\zeta\gtrsim
10^{-13}$ s$^{-1}$H$^{-1}$, but decreases almost to zero at
distances $\sim 0.1-0.3$ pc from the surfaces.

\section{Conclusion}

We showed in the paper that:
\begin{itemize}
\item From the HESS observations it follows that Sgr A$^\ast$ may be a source of high
energy protons which penetrate into molecular clouds. If the
injection spectrum of the protons continues into the MeV range,
then these protons generate continuous and line X-ray emission
from GC;
\item For the  derived injection rate of protons with energies
$\lesssim 100$ MeV the observed line and continuum emission from
hot plasma and molecular gas in the GC is generated by the same
protons;
\item
 Our model naturally explains the origin
of the two components  of nonthermal emission observed by {\it
Suzaku}, which are proportional to the 6.4 and 6.7 keV line
fluxes, respectively;
\item The observed  time-variations of the 6.4 keV flux from molecular clouds are strongly in
favor of the XRN model, but they do not exclude simultaneous
production of the 6.4 keV line by other processes, e.g., like this
one, especially as some observational results cannot be
interpreted in the framework of the XRN model;
\item It follows from our estimations that molecular clouds
may contribute a significant part of the total hard X-ray flux
from the GC.
\item
The width of the 6.4 keV line produced by protons is about several
tens of eV, which is about one order of magnitude wider than the
natural width expected from that generated by subrelativistic
electrons or X-ray reflection. Future observations by {\it Astro-H
SXS}, whose energy resolution is supposed to be only 7 eV
\citep{taka} will be able to measure this parameter and, thus, to
resolve the origin of the K$\alpha$ line emission from the GC;
\item Unlike other astrophysical problems, that of the 6.4 keV
flux can be solved in the near future because of its fast
time-variability. The 6.4 keV  flux from Sgr B2  had dropped for
the period from 2000 to 2005 to 60\% of its maximum value. If
 Sgr B2 is an XRN source then one expects  that this cloud will be almost
unseen in several years. If so, then no  explanation is acceptable
except the X-ray reflection. Otherwise, the origin of the emission
from Sgr B2 should be explained by other processes including this
one;
\item We showed that the density of subrelativistic protons is
almost constant in the GC, which explains the constant plasma
temperature in this region, $T\sim 6.5$ keV, as observed by {\it
Suzaku} and a constant surface brightness of molecular clouds in
the GC as observed by {\it SMM-Newton}, which cannot be
interpreted in the framework of the XRN model.

\end{itemize}

\vspace{5 mm} The authors thank the referee, K.Koyama, for very
useful advices, corrections and comments. The authors are also
grateful to F.Aharonian, A.Bykov, K.Kohri and H.Murakami for
discussions and to J.M.Nester for corrections of the text.

 VAD and DOC were partly supported by the RFBR
grant 08-02-00170-a, the NSC-RFBR Joint Research Project No
95WFA0700088 and by the grant of a President of the Russian
Federation "Scientific School of Academician V.L.Ginzburg". KSC is
supported by a RGC grant of the Hong Kong Government under HKU
7014/07P. CMK is supported in part by the National Science
Council, Taiwan under the grant NSC-96-2112-M-008-014-MY3.
A.~Bamba is supported by a JSPS Research Fellowship for Young
Scientists (19-1804).



\begin{thebibliography}{}
\bibitem[Aharonian et al (2008)]{aha}
Aharonian, F., Akhperjanian, A. G., Barres de Almeida, U. et al.
2008, A\&A, 483, 509
\bibitem[Bamba et al.(2002)]{bamba02}
Bamba, A., Murakami, H., Senda, A. et al. 2002, astro-ph/0202010
\bibitem[Bamba et al. (2009)]{bamba}
Bamba, A., Yamazaki, R., Kohri, K., Matsumoto, H., Wagner, S.,
P\"{u}hlhofer, G., \& Kosack, K. 2009, ApJ, 691, 1854

\bibitem[Berezinskii et al.(1990)]{ber90}
Berezinskii, V. S., Bulanov, S. V., Dogiel, V. A., Ginzburg, V.
L., and Ptuskin, V. S. 1990, {\it Astrophysics of Cosmic Rays},
ed. V.L.Ginzburg, (Norht-Holland, Amsterdam)

\bibitem[Cheng et al. (2007)]{cheng2}
Cheng, K. S.,  Chernyshov, D. O. \&  Dogiel, V. A. 2007, A\&A,
473, 351.

\bibitem[Dalgarno \& McCray(1972)]{dalmc}
Dalgarno A., \& McCray, R.A. 1972, ARA\&A, 10, 375
\bibitem[Dogiel et al.(1987)]{dog87}
Dogiel, V. A., Gurevich, A. V., Istomin, Ia. N., \& Zybin, K. P.
1987, MNRAS, 228, 843

\bibitem[Dogiel et al.(1998)]{dog4}
Dogiel, V. A., Ichimura, A., Inoue, H., \& Masai, K. 1998, PASJ,
50, 567

\bibitem[Dogiel et al.(2002)]{dog02}
Dogiel, V. A., Sch{\"o}nfelder, V., \& Strong, A. W.,  2002
 A\&A,  382, 730

\bibitem[Dogiel et al.(2009a)]{dog}
Dogiel, V. A., Tatischeff, V.,  Cheng, K-S., Chernyshov, D.O.,
 Ko, C.-M., \& Ip, W.-H., 2009a, A\&A, submitted
\bibitem[Dogiel et al.(2009b)]{dog_pasj}
Dogiel, V.,  Chernyshov D., Yuasa, T.,  Prokhorov, D., Cheng,
K.-S., Bamba, A., Inoue, H., Ko, C.-M., Kokubun, M.,  Maeda, Y.,
Mitsuda, K., Nakazawa, K., \& Yamasaki, N.Y.   2009b, PASJ,
submitted
\bibitem[Ebisawa et al.(2008)]{ebi2}
Ebisawa, K., Yamauchi, S., Tanaka, Y., Koyama, K., Ezoe, Y.,
Bamba, A., Kokubun, M., Hyodo, Y., Tsujimoto, M., \& Takahashi, H.
2008, PASJ, 60, 223
\bibitem[Garcia et al.(1973)]{garcia}
Garcia, J. D., Fortner, R. J., \& Kavanagh, T. M. 1973,
Rev.Mod.Phys., 45, 111
\bibitem[Goldsmith et al.(1990)]{lisg1}
Goldsmith, P. F., Lis, Dariusz C., Hills, R., \& Lasenby, J. 1990,
ApJ, 350, 186
\bibitem[Hayakawa et al.(1961)]{haya1}
Hayakawa, S., Nishimura, S., \& Takayanagi, K. 1961, PASJ, 13, 184
\bibitem[Hayakawa (1969)]{haya}
Hayakawa, S. 1969,  Cosmic Ray Physics (Wiley-Interscience)
\bibitem[Indriolo et al.(2009)]{indri}
Indriolo, N., Fields, B. D., \& McCall, B. J. 2009, to appear in
ApJ, astro-ph/0901.1143
\bibitem[Inui et al.(2009)]{inui}
Inui, T., Koyama, K., Matsumoto, H., \& Go Tsuru, T. 2009, PASJ,
61, S241
\bibitem[Koyama et al.(1996)]{koya1}
 Koyama, K., Maeda, Y., Sonobe, T., Takeshima, T., Tanaka, Y., \& Yamauchi, S.,
1996, \pasj, 48, 249
\bibitem[Koyama et al.(2007a)]{koya3}
Koyama, K., Inui, T., Hyodo, Y., Matsumoto, H., Go Tsuru, T.,
Maeda, Y., Murakami, H., Yamauchi, S., Kissel, S. E., Chan, K.-W.,
\& Soong, Y. 2007a, PASJ, 59, 221
\bibitem[Koyama et al.(2007b)]{koya2}
Koyama, K., Hyodo, Y., Inui, T. et al. 2007b, PASJ, 59, 245
\bibitem[Koyama et al.(2008a)]{koya4}
Koyama, K., Inui, T., Matsumoto, H., Go Tsuru, T.  2008a, PASJ,
60, 201

\bibitem[Koyama et al.(2009)]{koya09}
Koyama, K., Takikawa, Y., Hyodo, Y., Inui, T., Nobukawa, M.,
Matsumoto, H., \& Tsuru, T. G. 2009, PASJ, 61, S255

\bibitem[Lis \& Goldsmith(1989)]{lisg}
Lis, D. C., \& Goldsmith, P.F. 1989, ApJ, 337, 704
\bibitem[Morse \& Feshbach(1953)]{morse}
Morse, P. M., \& Feshbach, H. 1953, {\it Methods of theoretical
physics}, International Series in Pure and Applied Physics, New
York: McGraw-Hill
\bibitem[Muno et al.(2004)]{muno}
Muno, M. P., Baganoff, F. K., Bautz, M. W., Feigelson, E. D.,
Garmire, G. P., Morris, M. R., Park, S., Ricker, G. R., \&
Townsley, L. K. 2004, ApJ, 613, 326
\bibitem[Murakami et al.(2000)]{mura}
Murakami, H., Koyama, K., Sakano, M., Tsujimoto, M., \& Maeda, Y.
2000, ApJ, 534, 283
\bibitem[Murakami et al.(2001a)]{mura2}
Murakami, H., Koyama, K., Tsujimoto, M., Maeda, Y., \& Sakano, M.
2001a, ApJ, 550, 297
\bibitem[Murakami et al.(2001b)]{mura1}
Murakami, H., Koyama, K., \& Maeda, Y. 2001b, ApJ, 558, 687
\bibitem[Murakami et al.(2003)]{mura03}
Murakami, H., Senda, A., Maeda, Y., \& Koyama, K. 2003,
Astronomische Nachrichten, Supplementary Issue 1, Proceedings of
the Galactic Center Workshop 2002 - The central 300 parsecs of the
Milky Way., p.125
\bibitem[Nakajima et al.(2009)]{naka}
Nakajima, H., Go Tsuru, T., Nobukawa, M., Matsumoto, H., Koyama,
K., Murakami, H., Senda, A., \& Yamauchi, S. 2009, PASJ, 61, S233
\bibitem[Nath \& Biermann(1994)]{nath}
Nath, B.B., \& Biermann, P.L. 1994, MNRAS, 267, 447

\bibitem[Neufeld et al.(1995)]{neuf}
Neufeld, D. A., Lepp, S., \& Melnick, G. J. 1995, ApJS, 100, 132
\bibitem[Nobukawa et al.(2008)]{nobuk}
Nobukawa, M., Go Tsuru, T., Takikawa, Y. et al. 2008, PASJ, 60,
S191
\bibitem[Oka et al.(1998)]{oka}
Oka, T., Hasegawa, T., Hayashi, M., Handa, T., \& Sakamoto, S.
1998, ApJ, 493, 730
\bibitem[Predehl et al.(2003)]{predehl}
Predehl, P., Costantini, E., Hasinger, G., \& Tanaka, Y. 2003,
Astron.Nach., 324, 73
\bibitem[Revnivtsev et al.(2004)]{rev04}
Revnivtsev, M. G., Churazov, E. M., Sazonov, S. Yu. et al. 2004,
A\&A, 425, 49
\bibitem[Sacher \& Sch\"onfelder(1984)]{saher}
Sacher, W., \& Sch\"onfelder, V. 1984, ApJ, 279, 817
 \bibitem[Silk \& Werner(1969)]{siwer}
Silk J., \& Werner M. 1969, ApJ, 158, 185
\bibitem[Spitzer(1968)]{spt}
 Spitzer L. 1968,  Diffuse Matter in Space (Wiley Interscience, New York)
\bibitem[Spitzer \& Jenkins(1975)]{spj}
Spitzer, L., \& Jenkins, E.B. 1975, ARA\&A, 13, 133
\bibitem[Spitzer  \& Tomasko(1968)]{spi:68}
Spitzer, L., \& Tomasko, M.G. 1968, ApJ, 152, 971
\bibitem[Sunyaev et al.(1993)]{suny}
Sunyaev, R. A., Markevitch, M., \& Pavlinsky, M. 1993, ApJ, 407,
606
\bibitem[Sunyaev \& Churazov(1998)]{suny1}
Sunyaev, R., \& Churazov, E. 1998, MNRAS, 297, 1279
\bibitem[Syer \& Ulmer(1999)]{syer}
 Syer, D., \& Ulmer, A. 1999, MNRAS, 306, 35
\bibitem[Takahashi et al.(2008)]{taka}
Takahashi, T., Kelley, R., Mitsuda, K. et al. 2008,
astro-ph/0807.2007
\bibitem[Terrier et al.(2009)]{terr09}
Terrier, R., Belanger, G., Ponti, G., Trap, G., Goldwurm, A. \&
Decourchelle, A. 2009, Proc.of AIP, 2-d International Symbol-X
Symposium "Focusing on the Hard X-Ray Universe", submitted
 \bibitem[Valinia et al. (2000)]{val}
Valinia, A., Tatischeff, V., Arnaud, K., Ebisawa, K., \& Ramaty,
R. 2000, ApJ, 543, 733


\bibitem[Yuasa et al.(2008)]{yuasa}
Yuasa, T., Tamura, K., Nakazawa, K., Kokubun, M., Makishima, K.,
Bamba, A., Maeda, Y., Takahashi, T., Ebisawa, K., Senda, A.,
Hyodo, Y., Tsuru, T. G., Koyama, K., Yamauchi, S., \& Takahashi,
H. 2008, PASJ, 60, 207
\bibitem[Yusef-Zadeh et al.(2002)]{yus1}
Yusef-Zadeh, F., Law, C., \& Wardle, M. 2002, ApJ, 568, L121

\bibitem[Yusef-Zadeh et al.(2007a)]{yus2}
Yusef-Zadeh, F., Muno, M., Wardle, M., \& Lis, D. C. 2007a, ApJ,
656, 847

\bibitem[Yusef-Zadeh et al.(2007b)]{yus3}
Yusef-Zadeh, F., Wardle, M., \& Roy, S. 2007b, ApJ, 665, L123




\end{thebibliography}
\end{document}